\documentclass[onecolumn, pra]{revtex4}

\usepackage{epsfig}
\usepackage{amsfonts}
\usepackage{amssymb}
\usepackage{graphicx}
\usepackage{amsmath}

\usepackage{newlfont}
\usepackage{amssymb}
\usepackage{amsfonts}
\usepackage{amsmath}
\usepackage{wasysym}

\usepackage{bm}

\usepackage{amsthm}

\usepackage{color}
\usepackage{graphics} % eita aar graphicx clash kore gerakol na kore!

\usepackage{hyperref}
\hypersetup{colorlinks=true, urlcolor=blue}

\begin{document}

\renewcommand*{\DefineNamedColor}[4]{%
    \textcolor[named]{#2}{\rule{7mm}{7mm}}\quad
   \texttt{#2}\strut\\} 
% eita mone hoi dorkar nei!!

\definecolor{red}{rgb}{1,0,0}
\definecolor{green}{rgb}{0,1,0}
\definecolor{blue}{rgb}{0,0,1}
\definecolor{yellow}{rgb}{1,1,0}

\definecolor{cyan}{cmyk}{1,0,0,0} 

%cyan magenta yellow black

\definecolor{magenta}{cmyk}{0,1,0,0}

\definecolor{RHODAMINE}{cmyk}{0,0.82,0,0} %capital name use kora holo for ''section''!!!!!!!!!!!!!

\definecolor{Plum}{cmyk}{0.50,1.,0,0} % first letter-ta capital!!!!

\definecolor{OliveGreen}{cmyk}{0.64,0,0.95,0.40} % duto capital letter!!!!!

\definecolor{RedViolet}{cmyk}{0.07,0.90,0,0.34} % duto capital letter!!!!!

\definecolor{TealBlue}{cmyk}{0.86,0,0.34,0.02} % duto capital letter!!!!

\definecolor{BlueViolet}{cmyk}{0.86,0.91,0,0.04} % duto capital letter

\title{Quantum Advantage in Communication Networks}

\author{Aditi Sen(De) and Ujjwal Sen}

\begin{abstract}
 
Quantum channels are known to provide qualitatively better information transfer capacities over their classical counterparts. Examples include quantum cryptography,
quantum dense coding, and quantum teleportation. This is a short review on paradigmatic quantum communication protocols in both bipartite as well as multipartite 
scenarios.

\end{abstract}

\maketitle

\section{Introduction}

Communication is a process by which information is transferred from a sender  to a receiver. 
Information transmission networks have revolutionized our lives -- 
telephone, radio, television, internet, and other forms of communication, have become a necessity for most of us. 
Such transmission channels can be broadly classified into two categories -- ones without a security aspect, and ones with it.

Information transmission without a security feature can range from an examination result on the internet to a radio message from the meteorological office 
to warn about an impending storm or a telephone call from a relative.
%All the current classical communication schemes like internet, telephone etc. have revolutionized our day-to-day life.  
In such channels, there is either no a priori precaution taken to protect the message against potential eavesdroppers (as in telephone conversations)
or it is the intention of the sender that the message is obtained by as many receivers as possible (as in messages from the meteorological office). 
%
%
%%%%%%%%%%%%eta paltate hobe 
%However, we also need
However, one often requires
%%%%%%%%%%%%%%%%%
%At the same time, 
\emph{secure} communication protocols, and their applications range from day-to-day use e.g. in internet banking,  to national security (e.g. for 
communication between the security forces).
%have application ranging from  daily neccesity to the 
%security of the country. 

All the currently existing practical communication networks deal with classical messages sent over classical channels. By a classical message, we shall 
mean any information that can be expressed as an array of 0s and 1s, where  0 and  1 denote two distinguishable objects. Such distinguishable
objects can be a black ball and a white ball, 
or two distinguishable signals sent over an optical fibre. A classical channel is any transmission channel that carries such messages, and is
governed by the laws of classical mechanics. Mathematically, it is an operator that acts on arrays of 0s and 1s to produce similar arrays or 
probabilistic mixtures of them. 

Over the past few years, it has been realized that the performance of communication can be enhanced, sometimes even qualitatively, 
by using channels that observe quantum mechanical laws
%However, all the existing communication protocols  can be enhanced by using quantum mechanical laws, in particular, 
%by quantum correlations between two quantum mechanical subsystems of a  joint quantum state 
\cite{bbcrypto, Ekertcrypto, densec, tele}. After the theoretical 
proposals, the successful journey of quantum communication has begun from the experimental achievements starting in the late 1990s 
%\com{check the year!!!!} 
with photons \cite{phexp}. But several other physical systems including 
ions \cite{ionexp}, atoms in optical lattices \cite{atomsexp}, nuclear magnetic resonance \cite{nmrexp}, Josephson junctions \cite{ref-JJ}, and 
atoms and photons in a cavity \cite{ref-cavity},  have also been
used for implementing quantum information processing tasks.   While 
photonic devices and channels still remain
the trusted vehicle for long distance 
quantum communication, 
%monopoly
%\cite{long-distance}, 
experiments involving quantum communication via 
massive particles and those that involve atom-photon systems may turn out to be important for quantum computational applications.
% \cite{short-distance}. 

In 
%most of the 
quantum communication protocols that does not involve a security aspect, quantum correlations, aka entanglement \cite{reviewHHHH} has turned out to be an 
essential ingredient -- it is the \emph{resource} that runs the protocols. 
%useful resource. 
The quintessential quantum communication channels without a security feature are 
%bipartite (i.e. two-party) 
quantum states used as quantum dense coding \cite{densec} and quantum teleportation \cite{tele}
channels. They are channels respectively for transmitting
classical and quantum information, and form the basis of
most quantum channels without a security angle.
%The communication protocols (without involving the security issue) are of two types -- (1) classical information transmission via quantum states \cite{densec}, 
%and (2) quantum state  transfer using minimal amount of classical communication \cite{tele}. 
The advantage of both the protocols -- dense coding and teleportation -- over their classical counterparts,  
depend on the use of shared entangled quantum states. 
Before we present the communication schemes in further detail, we will give a formal definition of  entanglement in Sec. \ref{sec-ent}.

Classical information sent via quantum states will be discussed in Sec. \ref{sec-dense}. 
In 1992, C.H. Bennett and S.J. Wiesner invented a protocol -- called quantum dense coding \cite{densec} -- for sending classical messages by using a maximally 
entangled quantum state shared between a sender and a  receiver.
% and was called dense coding protocol. 
The information transmission capacity of this quantum channel is double of that of the corresponding classical channel.
Along with this initial protocol, we will also discuss the case when 
the shared quantum state may not be maximally entangled. 
%We will discuss a dense coding protocol which involve any 
%arbitrary state among the sender and the receiver along with the initial protocol of Bennett  and Wiesner. 
A communication channel with a 
%Since the 
single sender and a single receiver has 
limited commercial use, 
%In Section 2, 
and in Sec. \ref{sec-dense}, we  will  consider classical information transfer over quantum communication networks involving  
%generalization of such protocols for 
multiple senders and 
%multiple 
receivers.
% has immense importance in the communication world. 

While quantum dense coding involves sending classical information over a quantum channel, 
%Quantum state transmission using classical communication
``quantum teleportation'' \cite{tele} -- 
%was 
%first 
introduced in 1993 -- involves quantum information transfer over a quantum channel. 
%--
%  known as 
%quantum teleportation. 
It was shown that a  maximally entangled quantum state 
between the sender and the receiver, along with two bits of classical communication from the sender to the receiver,
%turns out to be maximized the rate of information transmission, 
is sufficient to exactly transfer the state of a two-dimensional quantum system (qubit). 
In the absence of the shared quantum state, the same transfer will take an infinite amount of classical communication. Therefore, while 
quantum dense coding results in a doubling  of the capacity of information transfer as compared to the corresponding classical protocol, 
quantum teleportation leads to an infinite resource reduction over its classical counterpart. 
These concepts will be discussed 
in Sec. \ref{sec-tele}. Other similar protocols involving multiple senders and multiple receivers will also be discussed in that section.

Sending secret messages has probably been 
%going on 
around
from the dawns of civilization. There has of course been huge advances in recent times, and e.g.
currently we use the internet for secure monetary transactions with a considerable amount of reliability. 
%eita paltate hobe!!!!!!!
However, all such practical classical 
cryptographic protocols depend for its security on unproven premises of the hardness of certain mathematical computations
on classical computers. [A classical computer is a device for performing mathematical operations, and which follows laws of classical mechanics 
for its functioning. Commercially available desktops, notebooks, etc. fall in this category. They are different from a quantum computer -- a device 
that performs mathematical operations, and follows laws of quantum mechanics for its functioning. Although a realistic quantum computer has as yet not been 
built, small-scale ones are being built in several laboratories around the globe. See e.g. \cite{lin-opt-qc, ionexp, atomsexp}] 
Most practical classical cryptographic schemes depend for their security 
on the unproven premise that it is hard to factorize an integer into its prime factors on a classical computer \cite{classical-crypto-book}. 
A twist to the story is that it is \emph{easy} to factorize an integer into its prime factors on a \emph{quantum} computer \cite{qc, easy-hard}. 
In 1984, in a conference of (mostly classical!) computer scientists in Bengaluru, C.H. Bennett and G. Brassard introduced a protocol 
%%Interestingly, while introducing 
 -- quantum cryptography \cite{bbcrypto} --
%quantum cryptography by C. H. Bennett and G. Brassarad  , 
for sending secret classical messages, where the security is guaranteed by the laws of quantum mechanics. In particular, such 
protocols remain secure even if the potential eavesdropper has a quantum computer to work with.   
Interestingly, the protocol did not require any shared entanglement. The existence of  
%only 
nonorthogonal quantum states in
%of 
quantum mechanics 
provided the security.
%was applied 
%
Later on, in 1991, A. Ekert proposed a quantum cryptographic scheme which involved shared entangled states \cite{Ekertcrypto}. 
From the perspective of security, both the protocols 
turn out to be equivalent \cite{device-independent}. We discuss about these protocols in Sec. \ref{sec-crypto}, along with generalizations to the case 
of multiple senders and receivers. In particular, we discuss a case of a single sender and two receivers where entanglement in the encoding states provide a 
higher level of security than in the case when there is no entanglement, thus demonstrating that entanglement is also an essential ingredient 
in \emph{secure} quantum communication.

\section{Entanglement}
\label{sec-ent}

Consider a situation where there are two observers
%
%Suppose there are two parties 
who are situated in two distant locations. It is usual to call them Alice and Bob. A cartoon of the situation is depicted in Fig. 1.
%\ref{ent-defn}.
\begin{figure}[h!]
\begin{center}
\label{ent-defn}
\epsfig{figure= 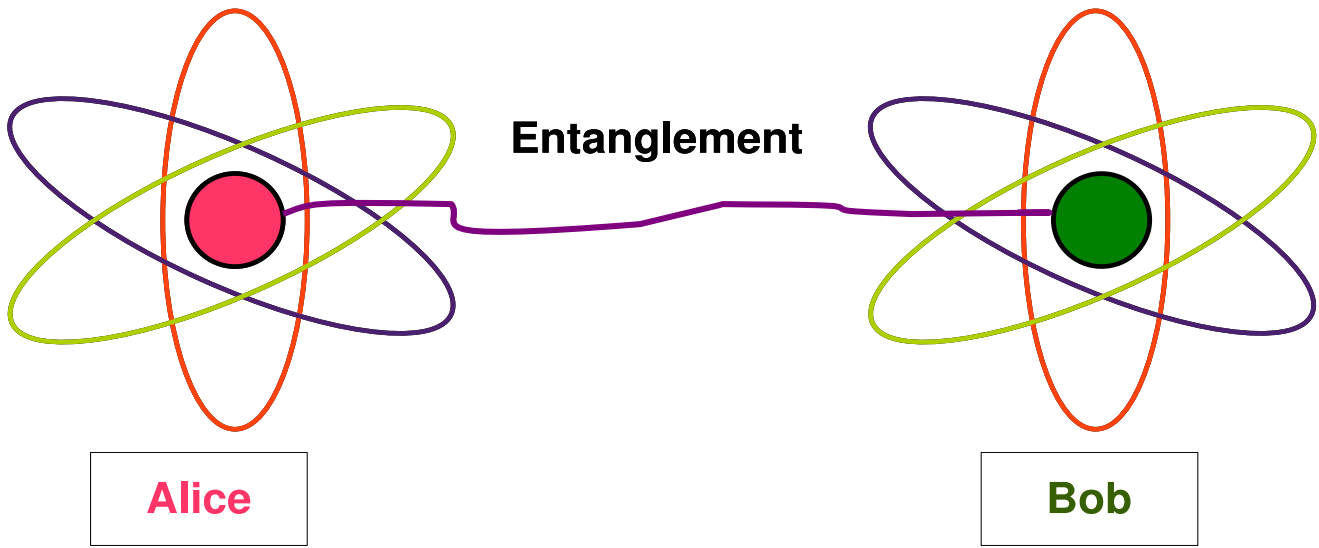, height=.55\textheight,width=1\textwidth}
\caption{Two separated observers -- Alice and Bob -- share an entangled state.
% The pictures of Alice and Bob were created at the website
%\texttt{http://morph.cs.st-andrews.ac.uk/Transformer/}. 
%Copyright not yet obtained!! Some other pics of Alice and Bob can also be given, if copyrights
%of those can be obtained.
%We thank Bernie Tiddeman, Department of Computer Science, Aberystwyth University, UK, for granting the approval.
}
\end{center}
\end{figure}
Below, unless mentioned to the contrary, 
a suffix \(A\) in the notation for a state or a Hilbert space
will imply that it is in possession of the observer Alice. And similarly \(B\) for Bob. 
Now suppose that Alice has a physical system, corresponding to which the quantum mechanical Hilbert space is \({\cal H}_A\), and prepares it in the quantum state
%prepares  
\(|\psi\rangle_A\).
% in \({\cal H}_A\) and 
Similarly, Bob prepares  the quantum state \(|\phi\rangle_B\) in \({\cal H}_B\). 
Then their joint state, \(|\Psi\rangle_{AB} = |\psi\rangle_A \otimes |\phi\rangle_B\), in \({\cal H}_A \otimes {\cal H}_B\),   
is said to be a (pure) product state.  
Note that the praparation of the joint state did not require any communication, classical or quantum, between the observers. 
Suppose now that the observers are allowed to communicate over a classical channel, say a phone line. In that case, the two observers can 
(classically) correlate their praparation procedures, and produce mixtures of product states, and the most general state that can be prapared 
in that way is of the form \cite{Werner89}
%Mathematically, it is given by
\[
\rho_{AB} = \sum_i p_i (|\psi_i\rangle \langle \psi_i|)_A \otimes (|\phi_i\rangle \langle \phi_i|)_B. 
%\rho_i^A \otimes \rho_i^B.
\] 
Here \(\{p_i\}\) forms a probability distribution, so that \(p_i \geq 0\), and \(\sum_i p_i =1\). 
This is therefore the most general shared quantum state of two parties that can be prepared by Alice and Bob, if they act locally in their distant laboratories
%In the case of density matrix, unentangled states or separable states can be created between Alice and Bob by using local quantum mechanical operations and 
%classical communication.
%The entangled states are those states which can be created by local operations and classical communication. 
with quantum operations and communicate over a classical channel. These set of operations is called ``LOCC'', standing for ``local (quantum) operations 
and classical communication.'' And the  bipartite (i.e. two-party) quantum states whose production is possible by LOCC are called 
separable states. Similar definitions are possible for multiparty (i.e. more than two parties) quantum states. 

Quantum states that cannot be prepared by LOCC are called entangled. Quantum theory allows the existence of 
such states (see \cite{reviewHHHH} for a review), and have actually been prepared in the lab (see e.g. \cite{phexp, ionexp, atomsexp}). 
%In other words, to create entanglement, two quantum mechanical particles should interact.
The preparation of such states necessarily requires an interaction between 
the two physical systems of Alice and Bob. 
%Pure entangled states are those states which can not be written in the form of \(|\psi\rangle_A \otimes \phi\rangle_B\) \cite{reviewHHHH}. 
 An example of an entangled state is the well-known singlet, 
%in \({\cal H^2}_A \otimes {\cal H^2}_B\) is 
\begin{equation}
\label{singlet}
 |\psi^{-} \rangle = \frac{1}{\sqrt{2}}(|01\rangle - |10\rangle),
\end{equation}
where \(|0\rangle\) and \(|1\rangle\) denote two orthonormal states of a qubit. They could e.g. be 
% respectively represent 
the spin-up and spin-down states in the \(z\)-direction of the spin degree of freedom of a spin-\(1/2\) particle, or 
the horizontal and vertical polarizations of a photon. As we will see in the subsequent sections,  the singlet state
% is known as
% singlet state and  
is a useful resource in quantum communication protocols. 
%The singlet state with other entnagled states have been  created  succesfully in 
%the laboratory with high visility.
For this and other reasons, the singlet state (and any other state that is local unitarily equivalent to it) is called a maximally entangled state. 
An equivalent definition of a maximally entangled (bipartite) state is a \emph{pure}
 bipartite quantum state whose local states are completely unpolarized -- a bipartite quantum state 
of which we have the complete information globally (as the von Neumann entropy of the total state is vanishing) 
while no information locally (as the von Neumann entropies of the local states are maximal).

\section{Classical Information Transmission via Quantum States: Dense Coding}
\label{sec-dense}

We will start by discussing the original quantum dense coding protocol \cite{densec} by using a maximally entangled state -- the singlet. 
Before proceeding to the quantum protocol, 
let us note that classically, to send two bits of classical information (i.e. four independent messages), one needs a four-dimensional system, i.e. 
four orthogonal states, which could e.g. be four differently coloured balls. 
%let us discuss a classical protocol and find
%estimate
% the maximum number of classical bits that can be sent using a  four-dimensional system via that protocol. Suppose 
%the sender (call Alice) wants to send four independent messages (say, \(A, B, C,  D\)), that is, \(2\) bits of classical information to the reciver (call Bob). It is 
%easy to see that Alice can encode \(2\) bits \cite{bit} in four orthogonal states, e.g. in four different color of balls and can send  one of them 
The protocol goes as follows. Suppose that Alice and Bob are together in Delhi today. Tomorrow, Alice will be in Mumbai, and Bob in Chennai. 
Suppose that after reaching Mumbai, Alice needs to send the information about the weather of the city to Bob. She is allowed to send 
only four independent messages (i.e. one option out of only four), and so while in Delhi they decide the following encoding: 
% to send depending on the message 
\begin{eqnarray}
\mbox{\textcolor{red}{Red} ball}  &:&     \mbox{Windy and raining}, \nonumber \\ 
\mbox{\textcolor{blue}{Blue} ball} &:&    \mbox{Windy but not raining}, \nonumber \\ 
\mbox{\textcolor{green}{Green} ball} &:&   \mbox{Not windy but raining}, \nonumber \\ 
\mbox{\textcolor{magenta}{Pink} ball} &:&  \mbox{Not windy and not raining.} \nonumber  
\end{eqnarray}
Since, the balls are distinguishable, Bob will be able to decode the message sent by Alice.  
%\begin{figure}[h!]
%\begin{center}
%\label{fig-mumbai-chennai}
%\epsfig{figure= chennai-mumbai.eps, height=.55\textheight,width=0.65\textwidth}
%\caption{It is about 1200 km by road from Mumbai to Chennai. \com{Map obtained from Google Maps. Copyright to be obtained before publication! Actually,
%a better pic should be given!!, which we were not able to obtain because of copyright concerns.}}
%\end{center}
%\end{figure}
%to Bob. Bob can decode the meassage as the encoding is in an  orthogonal states.  Classically, Alice requires \(4\)-dimensional system to send \(2\) bits of classical 
%information. 
The question is whether using quantum mechanics can help to increase the capacity of such classical information transmission. As we will now find out,
the answer is in the affirmative. 

Suppose that after Alice and Bob reach Mumbai and Chennai respectively, Alice creates a singlet in her lab in Mumbai, and sends half of it 
[i.e. one of the particles] to Bob. 
%[A singlet is composed of two qubits, and Alice sends one of them to Bob.] 
We assume that the resulting shared 
state between Alice and Bob is just the singlet state, so that the channel that carries half of the singlet from Alice 
to Bob is a noiseless quantum channel. This is currently science fiction, as the distance from Mumbai to Chennai is 
about a thousand kilometres,
% [cf. Fig. 2], 
and the distances between which entangled states can be created are currently about a hundred and fifty kilometres 
\cite{long-distance}. But 
not a long time ago, the same was around 10 kilometres \cite{10km}, and sometime before that, it was about 10 metres \cite{Aspect-er}!
%
%Alice and Bob subsytems can be described by the Hilbert spaces \({\cal H}_A\) and \({\cal H}_B\), the joint system of 
%Alice and Bob then is \({\cal H}_A \otimes {\cal H}_B\).   Now suppose Alice and Bob share a maximally entangled state (the singlet) in Eq. (\ref{singlet})
%\[
% |\psi^{-} \rangle = \frac{1}{\sqrt{2}}(|01\rangle - |10\rangle),
%\]
%where \(|0\rangle\) and \(|1\rangle\) respectively represent the up and down spins in the \(z\)-direction of a spin-\(1/2\) particle. 
After the singlet is created between Alice and Bob, 
%To send \(2\) bits of classical information, depending on the messages, 
Alice finds out the weather in Mumbai, and depending on what the weather is, she performs 
a unitary operation on her part of the shared singlet, according to the following instruction set: 
\begin{eqnarray}
 \mbox{Windy and raining} &:& \mathbb{I}, \nonumber \\ 
\mbox{Windy but not raining} &:& \sigma_z, \nonumber \\ 
\mbox{Not windy but raining} &:& \sigma_x, \nonumber \\ 
\mbox{Not windy and not raining} &:& \sigma_y, \nonumber  
\end{eqnarray}
where \(\mathbb{I}\) is the identity operation on the qubit Hilbert space, and \(\sigma_i\), \(i=x,y,z\) are the Pauli spin operators.
This has the following effect on the shared singlet:
%
%the following encoding with the help of the Pauli spin matrices 
%\(\{I, \sigma_i, i=x, y, z\}\), while Bob will do nothing on his part: 
\begin{eqnarray}
\label{eq-onek-holo-deri}
 \mathbb{I} \otimes \mathbb{I}  |\psi^{-} \rangle = |\psi^{-} \rangle = \frac{1}{\sqrt{2}}(|01\rangle - |10\rangle),\nonumber \\
\sigma_z \otimes \mathbb{I}   |\psi^{-} \rangle = |\psi^{+} \rangle = \frac{1}{\sqrt{2}}(|01\rangle + |10\rangle), \nonumber \\
 \sigma_x \otimes \mathbb{I}  |\psi^{-} \rangle = -|\phi^{-} \rangle = \frac{-1}{\sqrt{2}}(|00\rangle - |11\rangle), \nonumber \\
 \sigma_y \otimes \mathbb{I}  |\psi^{-} \rangle = \iota|\phi^{+} \rangle = \frac{\iota}{\sqrt{2}}(|00\rangle + |11\rangle), 
\end{eqnarray}
where \(\iota = \sqrt{-1}\), and where we suppose that in Eq. (\ref{eq-onek-holo-deri}), the first particle of the shared state is with Alice and the second 
is with Bob. Moreover, we assume that here and hereafter, 
\(|0\rangle\) and \(|1\rangle\) are eigenvectors of the Pauli \(\sigma_z\) operator, corresponding to the eigenvalues +1 and -1 respectively. 
 The states produced are either the singlet or the triplet states (perhaps up to a phase).
After the application of the unitary operation, according to the weather report obtained, Alice sends her part of the shared state to Bob, down a 
noiseless quantum channel, so that now the whole post-operated two-qubit state is with Bob. 
%The noiseless nature of the quantum channel is 
%not realistic, but we assume it  for mathematical simplicity.  
For the case of the quantum protocol, we assume that while in Delhi, Alice and Bob agree 
on the following encoding: 
\begin{eqnarray}
 \mbox{Windy and raining} &:& |\psi^{-} \rangle, \nonumber \\ 
\mbox{Windy but not raining} &:& |\psi^{+} \rangle, \nonumber\\ 
\mbox{Not windy but raining} &:& |\phi^{-} \rangle, \nonumber \\ 
\mbox{Not windy and not raining} &:& |\phi^{+} \rangle. \nonumber  
\end{eqnarray}
So e.g. if the state obtained by Bob is \(|\phi^+\rangle\) (up to an indeterminate phase), he will infer that the weather in Mumbai is ``not windy and not raining''. 
Since the singlet and the three triplets are mutually orthogonal, it is possible to set up a measurement to distinguish between them. Again this is 
not a simple matter to actually do that experiment in the lab \cite{Calsamiglia}, but is certainly a quantum mechanically allowed measurement. 
Therefore, after obtaining the second qubit from Alice, Bob measures the two-qubit state in the Bell basis (the two-qubit basis formed by the singlet and the 
three triplets), and finds out the weather in Mumbai, by looking up the encoding decided upon in Delhi.

Is there an advantage in the quantum protocol over the classical one? The classical protocol uses a four-dimensional physical 
system (a ball with four possible colours) to send two bits of classical information. In the quantum case, Alice initially sends a qubit (two-dimensional system)
to Bob to prepare the shared singlet, and subsequently sends another qubit after her unitary operations. So the total dimension of the physical system 
sent is again four \((=2 \times 2)\) in the quantum case, and again there are two bits of classical information sent from Alice to Bob. However, in the classical case, 
the whole four-dimensional state has to be sent after obtaining the news about the weather, while in the quantum protocol, the first two-dimensional system
can be sent before any news about the weather is obtained, and only the remaining two-dimensional system is to be sent after the news. It is in this sense 
that we have an advantage in the quantum case, and we say that the capacity of classical information transfer is doubled by using quantum mechanics --
two bits via a two-dimensional system in the quantum case against two bits via a four-dimensional system in the classical one, counting only the 
physical system sent after the news is obtained. One envisages a ``quantum internet'', where sending qubits is free at night, and costly during the day, so that 
Alice sends the first qubit at night to prepare the singlet, and sends only the second qubit during the day after the news about the weather is obtained. 
%This protocol shows that \(2\) dimensional system (i.e., the dimension of Alice's subsystem) is enough for encoding \(2\) bits of classical information, if Alice 
%and Bob primarily share an entangled state, while any classical protocol requires \(4\) dimensional system. 
While this may be still very much inside fiction, quantum dense coding (without such quantum memory effects) have been realized in the lab with 
several physical systems. See e.g. \cite{experimentsdense}.

 %\com{\textbf{experiments in dense coding}} \cite{experimentsdense}

\subsection{Towards Quantum Dense Coding Networks}

The dense coding protocol described before is for a maximally entangled state of two qubits. It can be easily generalized to the case 
of a maximally entangled state of higher dimensions, e.g.
\[
 |\Phi^+\rangle = \frac{1}{\sqrt{d}}\sum_{i=0}^{d-1} |ii\rangle,
\]
 in \({\cal H}_A \otimes {\cal H}_B\),
where \(d = \min\{d_A, d_B\}\), \(d_A = \dim {\cal H}_A\), \(d_B = \dim {\cal H}_B\), and where 
\(\{|i\rangle\}\) forms a mutually orthonormal set of states, to show that Alice can send \(\log_2 d_A + \log_2 d_B\) bits of classical information
by sending her part of 
\( |\Phi^+\rangle \) to Bob. 
Classically, by sending a \(d_A\)-dimensional system, Alice can send at most \(\log_2 d_A\) bits of classical information. 

Experiments often produce quantum states that are nonmaximally entangled and possibly noisy, and before we go over to networks, 
let us first find whether nonmaximally entangled states can also be used to obtain a quantum advantage 
in dense coding protocols. 

\subsubsection{Nonmaximally Entangled Bipartite States}

Consider therefore a bipartite quantum state (density matrix) \(\rho_{AB}\), defined on the tensor product Hilbert space
\({\cal H}_A \otimes {\cal H}_B\), and the question is
%
%However, producing an exact maximally entangled state \cite{experimentsdense} in the laboratory is restricted  due to the presence of different kind of noise. 
% In these limitations, it is important to know the capacity of classical information transmission while Alice and Bob share an arbitrary quantum state \(\rho_{AB}\) 
%in dimension \(d_A \otimes d_B\) (\(d_A\) and  \( d_B\) are respectively dimensions of Alice's Hilbert space \({\cal H}_A\) and  Bob's Hilbert space \({\cal H}_B\).  
%It can easily be checked that Alice can send only \(\log_2 (d_A)\) bits in the absence of an entangled state shared between Alice and Bob.  The original protocol shows 
%that (maximal) entanglement can enhance the capacity upto \(\log_2 d_A + \log_2 d_B\), provided there is a noiseless quantum channel betweenAlice and Bob. 
%
%
%We now discuss 
whether 
the capacity of classical information transfer, by using the shared quantum state \(\rho_{AB}\),
 can go beyond the classical limit of \(\log_2 d_A\).
%, when Alice and Bob share an arbitray quantum state 
%\(\rho_{AB}\). 
Suppose that the classical
information that Alice wants to send to Bob is \(i\), and that it happens with probability \(p_i\). 
As 
%described 
in the case of the original protocol of Bennett and Wiesner, Alice performs a unitary operation \(U_i\), if the message that she wants to send is 
\(i\),
% with probability
%\(p_i\) 
on her part of \(\rho_{AB}\).
%, depending on the message. 
Subsequent to
her unitary operation, she sends her part of the post-operated shared state
to Bob. Bob then has the ensemble \(\{p_i, \rho_i\}\), where
\[
\rho_i = U_i \otimes \mathbb{I}_{d_B} \rho_{AB} U_i^\dagger \otimes \mathbb{I}_{d_B},
\] 
with \(\mathbb{I}_{d_B}\) being the identity operator on Bob's physical system. 
The task of Bob now is to find as much information as is quantum mechanically possible about \(i\) from the ensemble 
\(\{p_i, \rho_i\}\). 
The procedure is schematically shown in Fig. 2. 
%\ref{fig-dense-flowchart}.
\begin{figure}[h!]
\begin{center}
\label{fig-dense-flowchart}
\epsfig{figure= 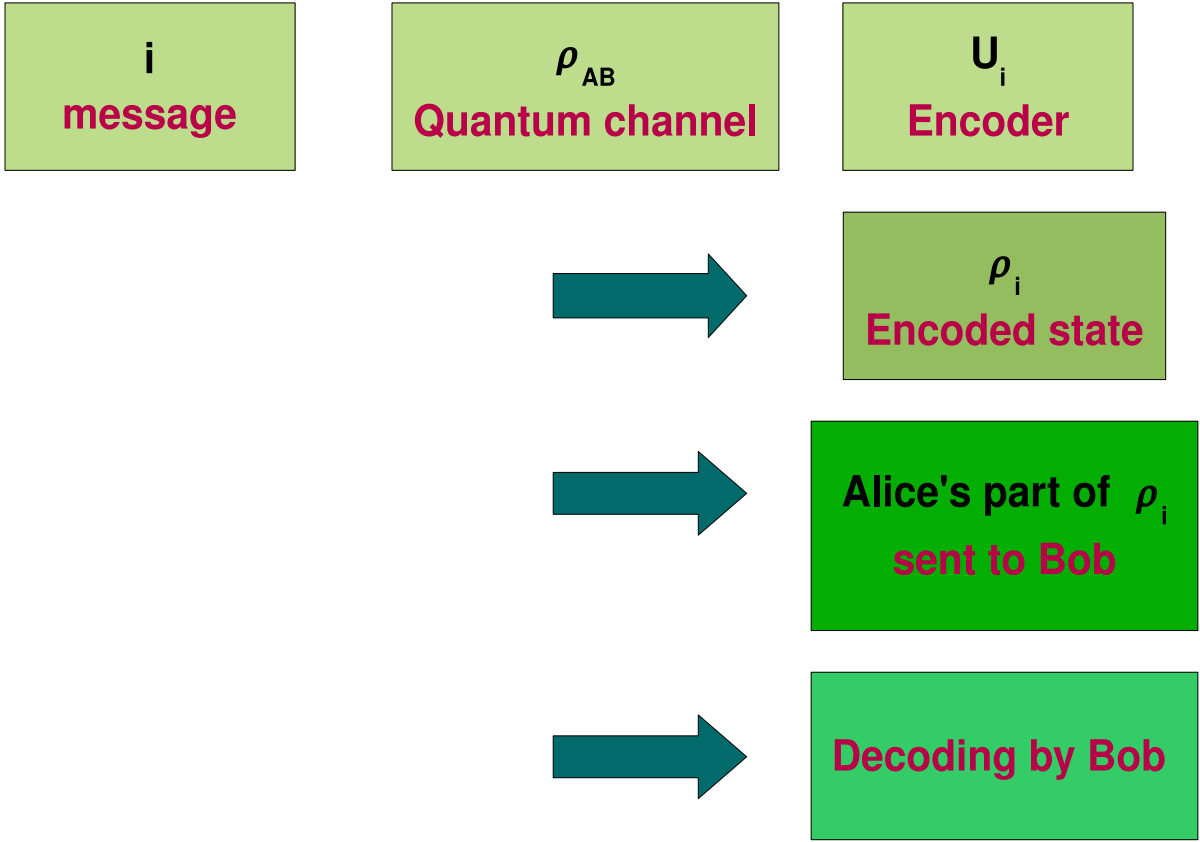, height=0.6\textheight,width=1\textwidth}
\caption{A flowchart for the quantum dense coding protocol.}
\end{center}
\end{figure}

%Schematically, therefore the procedure runs as follows:
%\begin{eqnarray}
% \underbrace{\quad \quad \quad i \quad \quad \quad}_{\mbox{message}} \quad \quad \quad \underbrace{\quad \quad \quad \rho_{AB} \quad \quad \quad}_{\mbox{quantum channel}}   
%\quad \quad \quad
%\underbrace{\quad \quad \quad U_i^A \quad \quad \quad}_{\mbox{encoder}} \nonumber \\ \nonumber \\
%\mapsto \underbrace{\quad \quad \quad \rho_i \quad \quad \quad}_{\mbox{encoded state}} \nonumber \\ \nonumber \\
%\mapsto \mbox{Alice's part of } \rho_i \mbox{ sent to Bob} \nonumber \\ \nonumber \\
%\mapsto \mbox{Decoding by Bob.} \nonumber
%\end{eqnarray}
%The suffix \(A\) on \(U_i\) indicates that the operation is performed by Alice. 

To decode the message, Bob performs a quantum  measurement on the ensemble  \(\{p_i, \rho_i\}\), 
 and obtains the result \(m\) with probability \(q_m\).
Let the corresponding post-measurement ensemble be \(\{p_{i|m},
\rho_{i|m}\}\). The information gathered by Bob (about the message \(i\)) can be quantified by the classical 
mutual information between  the message \(i\) and the
measurement outcome \(m\) \cite{CoverThomas}:
\begin{equation}
I(i:m)= H(\{p_i\}) - \sum_m q_m H(\{p_{i|m}\}).
\end{equation}
Here \(H(\{r_x\}) = -\sum_xr_x\log_2r_x\) is the Shannon entropy of the probability distribution \(\{r_x\}\). 
Note that \(H(\{p_i\})\) and  \(\sum_m q_m H(\{p_{i|m}\})\) respectively quantifies the pre-measurement and average post-measurement  ignorance (i.e., lack of information)
of Bob, about the message \(i\).  The difference \(I(i:m)\) therefore quantifies the information gain due to the measurement. 
%Note that the mutual information can be seen as the difference
%between the initial disorder and the (average) final disorder. Bob
%will be interested to obtain the maximal information, which is
%maximum of \(I(i:m)\) for all measurement strategies. 
%However, 
Bob's aim is to obtain as much information as possible about \(i\), and hence he has to perform a measurement that
maximizes  \(I(i:m)\) among  all quantum mechanically allowed  measurements. 
This leads us to the concept of accessible information,
\begin{equation}
I_{acc} = \max I(i:m),
\end{equation}
where the maximization is over all measurement strategies.
%  \(I_{acc}\) is known as accessible information. 

However, the maximization over measurements in accessible
information is in general hard to compute, and therefore it is important to obtain bounds on \(I_{acc}\) \cite{upper, lower, schumacher}.
% (cf.  \cite{schumacher, piotr, amader2}). 
%In this case, we will use the 
A universal upper bound  on \(I_{acc}\), called the ``Holevo bound'' \cite{upper}, 
is known for more than 30 years, and is
given by
\begin{equation}
I_{acc}(\{p_i, \rho_i\}) \leq \chi(\{p_i, \rho_i\}) \equiv
S(\overline{\rho}) - \sum_i p_i S(\rho_i),
\end{equation}
 where
\(\overline{\rho} = \sum_ip_i\rho_i\) is the average ensemble state,
and \(S(\varsigma)= - \mbox{tr} (\varsigma \log_2 \varsigma)\) is the
von Neumann entropy of \(\varsigma\).

Since the Holevo bound can be achieved asymptotically \cite{asymp, babarey, maarey, Hastings}, we define the capacity of dense coding  for the state \(\rho_{AB}\) as
\begin{equation}
\label{Kohinoor} 
{\cal C}(\rho) = 
%I_{acc}(\{p_i, \rho_i\}) \equiv 
\max_{p_i,U_i} \chi(\{p_i,
\rho_i\}) \equiv \max_{p_i,U_i} \left(S(\overline{\rho}) - \sum_i
p_i S(\rho_i)\right).
\end{equation}
The maximization involved in the capacity can be performed \cite{phoring,dui,dui0, dui1, tin,char} (see also \cite{ghasphoring}), and  we obtain  
\[
{\cal C}(\rho_{AB}) = \log_2 d_A + S(\rho_B) - S(\rho),
\]
where 
\(\rho_B = \mbox{tr}_A \rho_{AB}\) is Bob's part of the state \(\rho_{AB}\). 
The classical limit being at \(\log_2 d_A\), 
the bipartite quantum state \(\rho_{AB}\) is  useful for dense
coding if and only if  
\[
S_{A|B}(\rho_{AB}) =  S(\rho) - S(\rho_B)  < 0.
\] 
This reminds us of the classical conditional entropy (using Shannon, instead of von Neumann, entropy). While the classical expression 
is always positive, the corresponding quantum one (obtained by simply replacing the Shannon entropies by the von Neumann ones) can be negative, and 
exactly those bipartite quantum states which produce such a negative quantity are the ones which are useful for quantum dense coding
(cf. \cite{discord, work-deficit}, see also \cite{Michal-Jonathan-Andreas-Nature}).
\begin{figure}[h!]
\begin{center}
\label{fig-dimdense}
\epsfig{figure= 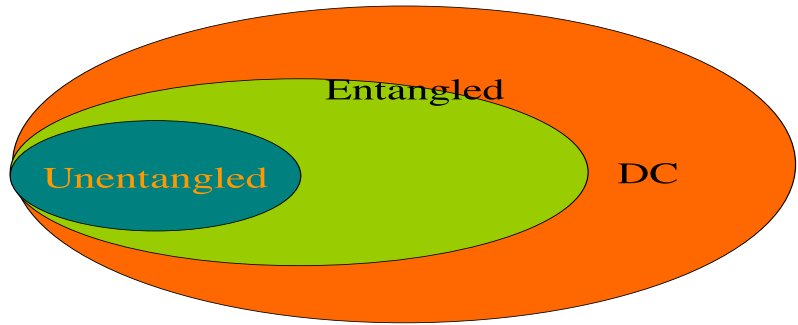, height=.25\textheight,width=0.65\textwidth}
\caption{Classification of bipartite
quantum states according to their usefulness in dense coding. The
%convex 
innermost region  consists of unentangled (separable) states.
The shell surrounding it contains states that are all entangled, but are not useful for dense coding. The outermost shell
is that of the dense-codeable states. It can be shown that both separable and non-DC states form convex sets \cite{Werner89, Wehrl}.}
\end{center}
\end{figure}
Such states will be called
dense-codeable (DC) states. DC
states exist, an example being the singlet state. It can be shown that no separable state can be used for dense coding \cite{tyaro}. 
Interestingly however, not all entangled states are useful for 
dense coding: there exist entangled states which have a positive \(S_{A|B}\) \cite{161718}. This leads us to a classification of 
%
%This criterion can be used to classify 
bipartite quantum states that is finer than that given by the entanglement-separability paradigm (see Fig. 3.)
%\ref{fig-dimdense}). 

\subsubsection{Beyond the Bipartite Setting: Networks}

Let us now briefly consider the relatively uncharted domain of quantum networks and their classical information transfer capabilities. 
To begin, let us consider a situation where we want to use an \(N\)-particle  
%Consider now a classical information transmission network  (via quantum channel). In the network, a 
quantum state \(\rho\) shared between 
\(N\) distant parties, with each party in possession of one particle in the beginning. 
Let us further suppose that we consider a situation where, 
of the \(N\) partners in the network,  \(N-1\) are to act as senders (call them \(A_1\), \(A_2\), \(\cdots\), \(A_{N-1}\), and there is 
a single receiver (call him Bob (\(B\))).
% \cite{char}. 
The senders want to send classical information to Bob by using the multiparty quantum state as a channel. 
It is possible to find the capacity of the multiparty quantum state for this classical communication task \cite{char}, 
%find Similar tool for the single sender and  the single receiver can be applied to obtain the capacity of dense coding 
and  is given by 
\[
{\cal C}_{\{A_i\} \rightarrow B}(\rho_{A_1 \cdots A_{N-1} B}) = 
%\underbrace{\log_2 d_{A_1} + \log_2 d_{A_2} +  \cdots + \log_2 d_{A_{N-1}}}_{\mbox{classical limit}} +  
{\cal C}_{\{A_i\} \rightarrow B}^{CL} + S(\rho_B) - S(\rho_{A_1  \ldots A_{N-1} B}),
\]
where the classical limit in this situation is given by 
\[
{\cal C}_{\{A_i\} \rightarrow B}^{CL} = \log_2 d_{A_1} + \log_2 d_{A_2} +  \cdots + \log_2 d_{A_{N-1}}.
 \]
In this case therefore, the state \(\rho\) is dense-codeable if and only if 
\[
  S(\rho_{A_1 \ldots A_{N-1} B})  -  S(\rho_B) < 0,
\]
with \(d_{A_i}\) being the dimension of the Hilbert space corresponding to the particle in possession of \(A_i\). 

%Due to t
The complexity 
of 
%mutipartite system, 
the  encoding-decoding processes 
%among many parties 
increases in the case when there are more than one receivers,
%become 
%intricated. Therefore, 
and 
the dense coding network 
between an arbitrary number of senders and an  arbitrary number of receivers is as yet not solved,
% problem, 
although some initial attempt was made  
to obtain the capacity 
%of dense coding 
in the case of an arbitrary number of senders and two receivers \cite{char, piotr}.

Further work on manipulation of classical information in multiparty quantum systems include Ref. \cite{Papadopoulos}, where 
%We establish 
a quantitative connection is established between the amount of lost classical information about a quantum state and the concomitant loss of entanglement, and
Ref. \cite{Borda},  where superadditivity of classical capacities of multi-access quantum channels is revealed.

\section{The Quantum No-cloning Theorem}
\label{sec-cloning}

Before proceeding further, we briefly discuss the no-cloning theorem in quantum mechanics. This will be important for our considerations in the succeeding 
sections, regarding quantum information transmission, as well as secret classical information transmission.

%An extreme case of gathering information about a state \(\left|\psi\right\rangle\) is 
%to prepare another copy of it while not disturbing the given copy. 
Suppose that a source produces the quantum states \(\left|\psi_0\right\rangle\) and  \(\left|\psi_1\right\rangle\).
% and 
%\((\left|0\right\rangle+ \left|1\right\rangle)/\sqrt{2}\). 
And our goal is to prepare two copies of the output.
Unitarity 
%The linearity 
of quantum mechanical evolutions makes such a goal impossible, unless \(\left|\psi_0\right\rangle\) and  \(\left|\psi_1\right\rangle\) are orthogonal \cite{nocloning}. 
More precisely, we want to have the transformations
\begin{equation}
\label{punching-lock-pack}
\left|\psi_0\right\rangle\left|B\right\rangle \left|M\right\rangle \rightarrow 
                                     \left|\psi_0\right\rangle\left|\psi_0\right\rangle \left|M_0\right\rangle \quad \mbox{and} \quad
\left|\psi_1\right\rangle\left|B\right\rangle \left|M\right\rangle \rightarrow 
                                \left|\psi_1\right\rangle\left|\psi_1\right\rangle\left|M_1\right\rangle,
\end{equation}
%and together, 
\emph{by the same quantum mechanical evolution}. 
%
%
%, we want 
%\[\frac{1}{\sqrt{2}}(\left|0\right\rangle + \left|1\right\rangle)\left|B\right\rangle 
 %                        																				\left|M\right\rangle \rightarrow 
%\frac{1}{\sqrt{2}}(\left|0\right\rangle + \left|1\right\rangle) 
%\frac{1}{\sqrt{2}}(\left|0\right\rangle + \left|1\right\rangle) \left|M_+\right\rangle.\]
Here \(\left|B\right\rangle\) is the blank state on which
the copy is to surface. And  \(\left|M\right\rangle\) is the initial state of the copying machine.
\(\left|M_0\right\rangle\) and \(\left|M_1\right\rangle\)
%   and     \(\left|M_+\right\rangle\)
are also machine states. So what we want is to ``clone'' the given state. 
Such machines are available to us in the classical world. The photo-copying machine in shops and offices does exactly this job: A paper containing whatsoever information 
is fed into the machine, along with a blank paper, and we get a copy of the original -- plus the original --
 after the job is completed, and the machine is maybe a bit heated-up
after the job. Such clones are also being tried with considerable success in the biological world. A parallel machine does not however exist in the 
quantum world. One may check that there is no single unitary operator by which both the transformations in (\ref{punching-lock-pack}) is possible, 
unless \(\left|\psi_0\right\rangle\) and  \(\left|\psi_1\right\rangle\) are orthogonal.

Although exact cloning is not possible
for arbitrary quantum states, it is possible to quantum mechanically clone 
a given set of states inexactly. This was first considered by 
Bu{\v z}ek and Hillery \cite{Buzek-Hillery-cloning}. Optimizations were considered by 
Bru{\ss} et al. \cite{Bruss-cloning}
(see also \cite{Gisin-MassarPRL-Werner-cloning-Keyl-Werner-cloning}). 
%See also \cite{arbitrary-blankPLA}.
Such inexact -- but near-optimal -- quantum cloners have actually been realized in the laboratory (see e.g. \cite{Boumeester}).

\textbf{No-information leaking}: Producing an exact clone of a state of a physical system is an extreme form of information gathering from the system. 
Is it possible to leak out a 
small amount of information about the state of a physical system, while still keeping the original state intact? 
Precisely, we want to have the transformations
\begin{equation}
\label{1998-e-proman-korechhilum-eita}
\left|\psi_0\right\rangle_S  \left|E\right\rangle_E \rightarrow 
                                     \left|\psi_0\right\rangle_S  \left|E_0\right\rangle_E \quad \mbox{and} \quad
\left|\psi_1\right\rangle_S  \left|E\right\rangle_E \rightarrow 
                                \left|\psi_1\right\rangle_S       \left|E_1\right\rangle_E,
\end{equation}
%and together, 
\emph{by the same quantum mechanical evolution}. Here the suffixes \(S\) and \(E\) denote ``system'' and ``environment''. If 
\(\left|E_0\right\rangle_E\) and \(\left|E_1\right\rangle_E\) are different for different 
\(\left|\psi_0\right\rangle_S\) and \(\left|\psi_1\right\rangle_S\), we will have leaked out information about whether the system is in 
\(\left|\psi_0\right\rangle_S\) or \(\left|\psi_1\right\rangle_S\), without disturbing (i.e. changing) the state of the system. 
However, it is easy to show that again such a transformation is not allowed in the quantum world, unless \(\left|\psi_0\right\rangle_S\) and \(\left|\psi_1\right\rangle_S\)
are orthogonal.

\section{Quantum State Transfer using Entangled States: Teleportation}
\label{sec-tele}

In Sec. \ref{sec-dense}, we considered transmission of classical information using quantum states. In this section,  we will consider transmission of quantum infomation 
via quantum states. Just as in Sec. \ref{sec-dense}, we will begin with the case when there is a single sender and a single receiver.
The task of the sender, Alice,  is to send an arbitrary quantum state to a distant receiver, Bob, 
with whom she has a previously shared bipartite quantum state. We begin with the case when this shared state is the singlet and when the 
sent quantum state is a qubit, as 
was done in the first paper that considered the protocol. The protocol was called quantum teleportation \cite{tele}. 
In addition to the shared bipartite quantum state that is to be used by Alice and Bob as the resource for sending the arbitrary qubit, 
Alice and Bob are also allowed to use a limited amount of classical communication. 

If the qubit is in a known state (known only to the sender, and not the receiver), 
then the sender may try to send that knowledge to the receiver, classically, say over a phone line. 
However, this requires an infinite number of bits 
of classical data transfer, as an arbitrary (pure) qubit can be written as \(\cos(\theta/2) |0\rangle + \exp(\phi) \sin (\theta/2)|1\rangle\) 
(\(\theta \in [0, \pi]\), \(\phi \in [0, 2\pi)\)), which can be represented as a point on a unit sphere, the ``Bloch'' or the ``Poincar{\'e}'' sphere. 
For an arbitrary qubit, 
Alice
%If Alice knows the qubit, she 
needs to send the information about \(\theta\) and \(\phi\) to Bob, and sending it classically requires an infinite amount of 
data transfer, as the range of these parameters are continuous infinities. Sending the information over a classical channel is not an option in the case when
the qubit is in an unknown state and only a single copy is provided, as it is not possible to find the state of a qubit from a single copy. Since an
unknown  qubit cannot be cloned, as we saw in the preceeding section, it is not possible to produce many copies from the given single copy.

Readers should notice that in quantum  state transmission, we use a different set of resources than in 
classical information transmission, discussed in Sec. \ref{sec-dense}. 
In the latter case, considering only the situation of a single sender and a single receiver (the other cases being similar), 
the resources were the entangled shared state between the sender and the receiver, and the noiseless quantum channel from the sender to the receiver. 
In particular, classical communication, say over a phone line, between the sender and the receiver was forbidden: Sending classical communication is 
the task of the protocol.
In the 
quantum case, the resources are 
the entangled shared state between the sender and the receiver, and a limited amount of classical communication. Quantum communication -- say over a 
noiseless quantum channel -- is forbidden: Sending quantum states is in this case the task of the protocol. 
%the sender's aim is to minimize the classical communication between the sender and the receiver. 

We now describe the  protocol of quantum mechanically teleporting a qubit from Alice to Bob, where the previously shared quantum state is the singlet.
%
%Suppose Alice and Bob share a maximally entangled state \(|\psi^{-} \rangle = \frac{1}{\sqrt{2}}(|01\rangle - |10\rangle)\), same as in Eq. (\ref{singlet}). 
Alice wants to send an arbitrary quantum state \(|\chi \rangle = a |0\rangle + b|1\rangle\), where \(a\) and \(b\) are complex numbers with the 
normalization \(|a|^2 + |b|^2 =1\). 
To send the state, Alice performs a measurement in the Bell basis, \(\{|\psi^{\pm} \rangle, |\phi^{\pm} \rangle\}\), on the four-dimensional 
space formed by her part of the singlet and the input state \(|\chi\rangle\). 
%She communicates 
%the measurement result to Bob. 
Depending on the measurement results, there are different states created at Bob's end, as shown in the following table.
\begin{table}
\label{tele_table11}
\centering
\begin{tabular}{|c|c|}
\hline
Measurement outcome at Alice & State created at Bob \\
%j & \ket{\psi^{j,0}} & \ket{\psi^{j,1}} & \sigma_1^{j,k} \otimes \sigma_2^{j,k} \\
\hline
 \(|\psi^{-}\rangle\) & \(\phantom{\sigma_z}|\chi\rangle\) \\
        \(    |\psi^{+} \rangle\) & \(\sigma_z |\chi\rangle\) \\
         \(   |\phi^{-} \rangle\) & \(\sigma_x   |\chi\rangle\) \\
           \( |\phi^{+} \rangle \)&  \(\sigma_y |\chi\rangle\) \\
            \hline
%\mbox{Table 1.} 
\end{tabular}
\caption{Correlations between measurement results at Alice and the corresponding states at Bob.}
%\nonumber
\end{table}
This can be easily checked by explicit calculation. For example, if the outcome at Alice's end is \(|\phi^-\rangle\), the state created at Bob is 
\(\left(\langle \phi^-| \right)_{1A} \left(|\chi\rangle_1 \otimes |\psi^-\rangle_{AB} \right)\), which is exactly \(\sigma_x   |\chi\rangle\), upto a 
constant multiple. We have ignored multiplicative phases in the above table.  We use the suffix ``1'' to denote the input qubit.

Alice communicates the result of the measurement to Bob over a classical channel. Note that the number of bits that needs to be sent is 2 (corresponding to four outcomes). 

Depending on the message received by Bob, 
%measurement results (see Table \ref{tele_table}), 
he performs  one of the unitaries \(\{\mathbb{I}, \sigma_i, i =x, y, z\}\)  
on his particle according to the following instruction set.
% and obtains the state \(|\chi\rangle\). The protocol is depicted in Fig. \ref{fig-tele}.
\begin{equation}
%\label{tele_table}
\begin{array}{|c|c|}
\hline
\mbox{Measurement outcome at Alice} & \mbox{Unitary applied by Bob} \\
%j & \ket{\psi^{j,0}} & \ket{\psi^{j,1}} & \sigma_1^{j,k} \otimes \sigma_2^{j,k} \\
\hline
 |\psi^{-}\rangle & \mathbb{I} \\
            |\psi^{+} \rangle &\sigma_z \\
            |\phi^{-} \rangle &\sigma_x  \\
            |\phi^{+} \rangle & \sigma_y\\
            \hline
%\mbox{Table 1.} 
\end{array}
\nonumber
\end{equation}
After performing the unitary operation, Bob's state is exactly in the state \(|\chi\rangle\), the state used by Alice as the initial state. 
We have therefore been able to use an entangled quantum state (the singlet) to send an arbitrary qubit from Alice to Bob, 
without actually sending the qubit down a quantum channel.
The protocol is pictorially depicted in Fig. 4. 
%\ref{fig-tele}.
%
%
%
%
\begin{figure}[h!]
\begin{center}
\label{fig-tele}
\epsfig{figure= 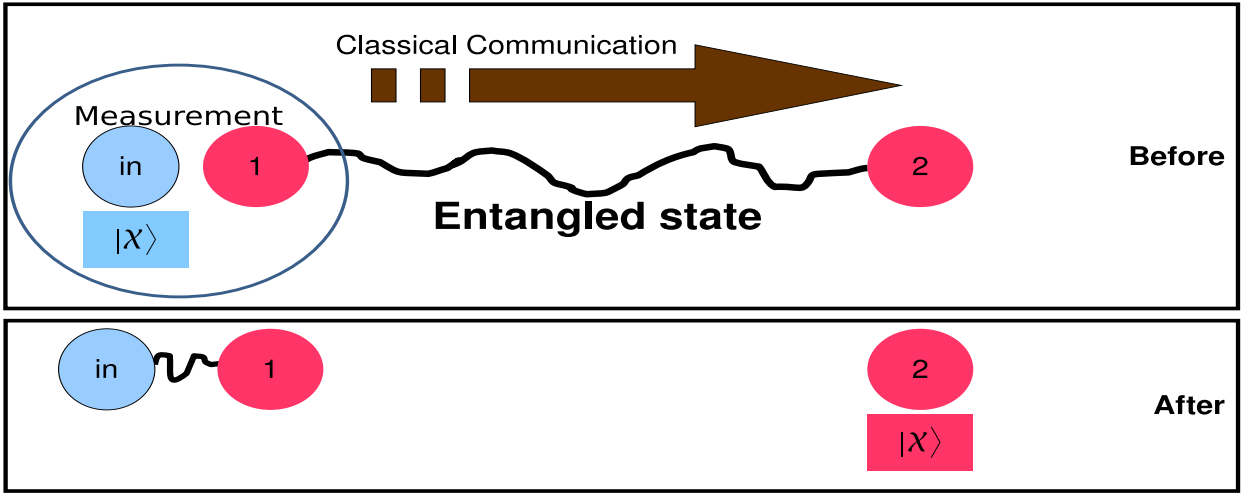, height=.48\textheight,width=1\textwidth}
\caption{A pictorial description of the teleportation protocol.}
\end{center}
\end{figure}

Can one use the quantum teleportation protocol for (superluminal) signaling? In the protocol as given above, Alice sends the measurement outcome 
to Bob over a classical channel, which will of course respect special relativity. But what if Alice does not send her message? The states in the right
column of Table 1 
%\ref{tele_table11} 
are still created at Bob's end. However, without Alice's message, Bob only knows that a measurement onto the basis 
\(\{|\psi^{\pm} \rangle, |\phi^{\pm} \rangle\}\) is performed by Alice, and that any one of the states  \(\{|\psi^{\pm} \rangle, |\phi^{\pm} \rangle\}\) are 
obtained by her -- he does not know which! To find his post-measurement state, Bob must therefore consider a mixture of the states in 
the right
column of Table 1, 
%\ref{tele_table11}, 
with the corresponding probabilities -- which he can find out by using Born rule. A simple calculation shows that 
the resulting mixture is the completely unpolarized state, and is equal to his pre-measurement state \(\mbox{tr}_A(|\psi^-\rangle \langle\psi^-|)\). 
Quantum teleportation therefore does not help us to signal.

The second question is whether one can use quantum teleportation for violating the no-cloning theorem. The answer is again negative. 
Although an exact copy of the input state is created 
at Bob's end, the original copy is completely destroyed. The post-measurement state of Alice does not contain any information whatsoever about the input state 
\(|\chi\rangle\): The output states at Alice are the states of the Bell basis and are independent of 
\(a\) and \(b\). So are the probabilities of their occurrence.

%GENERAL KICHHU LIKHTE HOBE

Experimental realization of quantum teleportation has been achieved in many physical systems, like photons, ions, and light-matter interface. See e.g. 
\cite{tele-expt}.

\subsection{Teleportation with Nonmaximally Entangled Bipartite States}

A shared singlet state, along with limited classical communication, supports an exact transfer of a qubit. Nonmaximally entangled states, however, 
do not enjoy such a feature. To go further, we have to introduce the concept of ``fidelity''. Suppose that a machine promises to produce a quantum state 
\(|\psi\rangle\). However, it actually produces the state \(\zeta\), which 
may be mixed \cite{mixed-input}. We then say that the machine has the fidelity \(\langle \psi| \zeta |\psi \rangle\). In case the machine is 
exact, the fidelity is unity. 
In connection to quantum teleportation, suppose that Alice and Bob share the quantum state \(\rho_{AB}\), defined on the Hilbert space
\({\cal H}_A \otimes {\cal H}_B\). For simplicity, we assume that \(\dim {\cal H}_A = \dim {\cal H}_B = d\). The task now is to transfer the
state \(|\psi\rangle\) of a \(d\)-dimensional 
quantum system from Alice to Bob by using the quantum state \(\rho_{AB}\) as a quantum channel. Suppose that the state produced at Bob's end 
after the teleportation protocol is \(\zeta (\psi)\). Then the fidelity of the teleportation protocol is 
\[
 F = \int \langle \psi| \zeta(\psi) | \psi \rangle d\psi \Big/ \int d\psi,
\]
where we have included a uniform average over the \(d\)-dimensional Hilbert space of the input state.

If Alice and Bob do not share the quantum state \(\rho\), but Alice still wants to send the unknown quantum system \(|\psi\rangle\) -- of which she 
has a single copy-- to Bob by sending  classical 
information over a classical channel, the fidelity that she can reach is just \(2/(d+1)\). The optimal protocol for Alice to follow is to 
measure the quantum state \(|\psi\rangle\) in an arbitrary basis, and send the measurement outcome to Bob, after which Bob prepares a quantum system in that 
basis state. 
Let us call the limit \(2/(d+1)\) as the classical fidelity \cite{tele-HHH}.

There do exist nonmaximally entangled states that support quantum teleportation with a fidelity that is higher than the classical fidelity. An 
important result was obtained by the Horodeccy family, where they found a relation between the optimal teleporation fidelity
and the singlet fraction (fidelity of a two-party state to the corresponding maximally entangled state) 
 \cite{tele-HHH} (see also \cite{Nielsen-tele}), and ruled out the ability of a large class entangled quantum states to teleport 
with fidelity beyond the classical limit.

\subsection{Quantum Networks carrying Quantum Information}

There are many interesting examples of quantum information transmission in quantum networks, and below we provide a necessarily incomplete selection. 
The subject is replete with open problems.

\subsubsection{Entanglement Swapping}

Close on the heels of the teleportation paper, came the paper on entanglement swapping \cite{ent-swap} (see also \cite{ent-swap+}), which, along with 
giving an interesting example of quantum information transmission and a proposal for preparation of bipartite entangled states, also 
led to a re-thinking of the definition of entangled states. Let us first describe the protocol. Suppose that two distant  observers Alice and Bob 
share the singlet state. Two other observers, Charu and Debu,  share another singlet. We assume that Alice and Debu have never met in the past and neither 
do they have a quantum channel between them -- i.e. their particles
have never interacted. The particles with Alice and Bob have of course interacted in the past, as otherwise they could not have been in the 
singlet state. Also Charu's and Debu's particles must have interacted in the past. Now suppose that Bob and Charu meet and perform a measurement 
%(in the Bell 
%basis) 
on the particles that they carry. 
%See Fig. \com{ekhane ekta fig dewa hobe ki?}. 
It turns out that it is now possible for Alice and Debu -- whose particles have 
never interacted in the past -- to share an entangled state. 
There have been several experiments that have realized entanglement swapping
\cite{swap-expt}. 
The concept has been found to be useful e.g. for proposals to realize 
long distance quantum communication via ``quantum repeaters'' (see e.g. \cite{repeaters, output-state}).

\subsubsection{Quantum Telecloning}

Quantum telecloning is 
%Murao \emph{et al.} \cite{telecloning} consider the 
a  natural generalization of the original teleportation protocol 
%involving a single sender and a single receiver 
to the case where there is still a single sender, but many receivers \cite{telecloning}. 
[A similar protocol was previously considered in Ref. \cite{Bruss-cloning}.]
Suppose that Alice is in possession of a 
qubit in an unknown  state \(|\psi\rangle\). She wishes to send it to \(N-1\) distant receivers \(B_1\), \(B_2\), ...,  \(B_{N-1}\) (\(N>2\)). 
Producing exact copies at all these locations is forbidden by the no-cloning theorem. However, Alice may try to prepare \(N-1\) inexact 
copies of \(|\psi\rangle\) and send it to the \(N-1\) distant locations by using \(N-1\) singlets which she happens to share with the \(N-1\) 
distant receivers. Murao \emph{et al.} showed that the same result can be obtained if Alice and her \(N-1\) associates share particular multipartite 
entangled states, in which case only \(O(\log_2(N-1))\) singlets are required. 
The protocol has been experimentally realized -- see \cite{telecloning-expt}, and references therein.

\subsubsection{Multi-access Channels and Multiparty Entanglement}

For 
%higher entanglement for a 
a pure \emph{bipartite} quantum state, a higher entanglement 
implies higher capacities for sending both classical and quantum information by using the bipartite state as a quantum channel \cite{densec, tele, pure-ent,
phoring, dui, dui1,  Shor-capa, tin, char}. 
%instances in the case of a single sender and a single receiver.
This relatively simple image in the bipartite domain is not mirrored in the case of many senders and many receivers. 
More precisely, two types of multi-access quantum channels, motivated by distillation protocols in multiparty quantum
networks, were defined in Ref. \cite{multi-access-amader}, and it was shown that the capacities of these channels 
%capacities of four-party quantum states that are motivated
%by considering distillation protocols in multiparty quantum
%networks and show that their values are 
cannot be correlated with \emph{any} multiparty entanglement measure.
%those of a measure of genuine four-party entanglement.

\section{Quantum Cryptography} 
\label{sec-crypto}

%In Section 2, we will discuss a protocol of classical information transmission via quantum states \cite{densec}.

Until now, we have discussed about communication protocols that do not have a security aspect. 
In this section, we will deal with protocols for sending \emph{secret} classical information over a quantum channel. 
It is worthwhile to mention here that among the exciting achievements in quantum information science during the last few years, 
the experimental success of quantum cryptography 
 is one of 
the foremost
%them. For details, see 
\cite{reviewcrypto}. 

A typical cryptographic scheme consists of 
a sender, Alice, who wants to send some classical message to a receiver, Bob, secretly. 
Since all messages can be written as a sequence of binary digits (0's and 1's), we suppose that Alice has a 
sequence \(P\) of 0's and 1's which she wants to send to Bob. This is known in cryptographic parlance as ``plaintext''.
If we now suppose that Alice and Bob have a previously shared ``secret key'' \(K\) -- a random sequence of 0's and 1's that both Alice and Bob know, but nobody else knows 
it -- which is at least as long as the plaintext, then Alice will be able to send her plaintext secretly to Bob, by a protocol known as the ``one-time pad''. 
The protocol can be described as 
follows. Alice adds (bitwise addition modulo 2) \(P\) and \(K\) to obtain the ``ciphertext'' \(C\), and sends it down a classical channel to Bob. 
Since \(K\) is random, \(C\) is also random, and hence it is not possible 
to get any information from it. So, even if an eavesdropper (it is usual to call her ``Eve'') gets hold of \(C\), she is 
not able to decipher it.  However, once it reaches Bob, he again adds \(K\) to \(C\), to obtain the original plaintext \(P\). 
%Let us begin by describing There exists an efficient classical cryptography protocol, known as one-time pad which 
%can send the message securely. 
As an illustration, suppose that 
% an example of a  sending protocol by one-time pad. Suppose 
Alice wants to send some message which is   \( 01010110\) -- the plaintext.
%(call Plaintext)  after encoding in a binary digits. 
Moreover, Alice has a random sequence of binary digits, \(00111001\) -- the key. 
Bob has exactly the same random sequence, but nobody else has it. 
%share  a completely random  but an equivalent key \(00111001\). Therefore, the encoding  of Alice is as follows:
Alice encodes her message by adding the message to the key:
\begin{eqnarray}
  \mbox{ Plaintext }  +  \mbox{ key }  (\mbox{modulo } 2) &=& \mbox{ ciphertext}, \nonumber\\
\mbox{i.e. }   01010110 +  00111001 \quad (\mbox{modulo } 2) &=& 01101111. \nonumber
\end{eqnarray}
After the binary addition, Alice sends the ciphertext to Bob, who after obtaining it goes through the decoding process: 
%After the message has been sent, Bob's decoding process is
\begin{eqnarray}
  \mbox{ Ciphertext }  +  \mbox{ key }  (\mbox{modulo } 2) &=& \mbox{ plaintext}, \nonumber\\
\mbox{i.e. } 01101111   +  00111001 \quad (\mbox{modulo } 2) &=& 01010110. \nonumber
\end{eqnarray}
%
%
%\[
% \mbox{Decoding: Plaintext}   = \mbox{Ciphertext} + \mbox{key} (\mbox{modulo} 2)\]
This process of sending a secret message is completely secure, provided the key is never re-used. 
Therefore, having shared secret keys is sufficient to implement cryptography.
%However, 
%, provided the  key is as long as the message and the key is completely random.
%Of course, the key is not known to anyone else. Reusing the same key can  abort the security of the method. 
%However, the security of this protocol depends on the fact that there exist a method by which 
% Alice and Bob can share a exactly same  random binary sequence,  and no-one else has it.
Generation of such  secret keys  
is trivially possible 
if Alice and Bob meet before they actually send the encoded message.
The question is whether they can generate secret keys without meeting.

The cryptography problem therefore boils down to  a key  distribution protocol.
Generation of such secret keys, without meeting, is possible both in the classical and the quantum worlds. In both cases, the security depends 
on some unproven premises.   
%Such key generation is also possible without using quantum mechanics. 
In the classical case, the security of practical key distribution protocols 
%security 
depends on the unproven premise of the hardness in solving 
certain mathematical equations. Discussion of such classical protocols is beyond the scope of this review \cite{classical-crypto-book}. 
%We will not discuss these protocols here. 
%On the other hand, i
In the quantum case, 
the security of the quantum key distribution protocols depends on the validity of quantum physics. 

We will now discuss the protocol of quantum key distribution known as the Bennett-Brassard 1984 (BB84) protocol \cite{bbcrypto}.  
Of the striking facts of the protocol is that it does not involve shared entangled states between the sender and the receiver.  
%cryptography protocol.
To begin, the sender,  Alice,  randomly chooses one of the following two bases: 
%a prepares a qubit in a state from any one one of the Bases
\begin{eqnarray}
\mbox{Basis } Z &=& \{|0\rangle,|1\rangle\},  \nonumber\\
\mbox{Basis } X &=& \Big\{\frac{1}{\sqrt{2}}(|0\rangle+|1\rangle), \frac{1}{\sqrt{2}}(|0\rangle-|1\rangle)\Big\}.
\end{eqnarray}
%where \(|\pm z\rangle\) and \(|\pm x\rangle\) 
As the names suggest, they are formed by respectively the eigenvectors of the Pauli spin matrices \(\sigma_z\) and \(\sigma_x\).
After choosing the basis, she randomly chooses a state from that basis, prepares a qubit in that state, 
 and sends the qubit to Bob over a quantum channel.
After obtaining the state, Bob randomly measures in basis \(Z\)  or basis \(X\). 
Alice and Bob then publicly declare (e.g. by advertising on a newspaper) their bases -- Alice conveys the basis that she chose for preparing the qubit, and 
Bob discloses the basis in which he measures. 
They certainly do \emph{not} declare the states -- Alice does not announce the state she prepared, and Bob does not 
reveal the state he obtained as the measurement outcome. 
They repeat this procedure over many runs. For each run, Alice notes down a 0 if she had chosen an eigenvector corresponding to the eigenvalue +1, i.e.
if she had chosen either 
\(|0\rangle\) of basis \(Z\) or \(\frac{1}{2}(|0\rangle+|1\rangle)\) of basis \(X\). She notes down a 1 otherwise. Similarly, 
if Bob measures in the basis \(Z\), and gets the outcome 
\(|0\rangle\) (therefore corresponding to eigenvalue +1 of \(\sigma_z\)) or if he 
measures in the basis \(X\), and gets the outcome \(\frac{1}{2}(|0\rangle+|1\rangle)\)
 (again corresponding to eigenvalue +1, but of \(\sigma_x\)), he notes down a 0. He notes down a 1 otherwise. In this way, they obtain a 
sequence of 0's and 1's. Due to the preparation procedure, at least the sequence at Alice is a random one. The one at Bob may depend on the noise 
in the quantum channel, which may in principle be due to an eavesdropper who is trying to get information about the sent qubits. 
After the bases are publicly declared, Alice and Bob remove the entries in their corresponding binary sequences that correspond to runs for which 
the bases did not match.

At this stage, if there is no noise in the quantum channel, the binary sequences at Alice and Bob should exactly match. Let us 
consider a few exemplary runs, where for simplicity we suppose that there is no noise in the channel. 
\begin{equation}
\label{crypto_table}
\begin{array}{|c|c|c|c|c|c|}
\hline
                         & \mbox{Alice's state}                   & \mbox{Alice's bit}  & \mbox{Bob's basis} & \mbox{Bob's bit} & \mbox{Decision}       \\
%\mbox{Measurement outcome at Alice} & \mbox{Unitary applied by Bob} \\
%j & \ket{\psi^{j,0}} & \ket{\psi^{j,1}} & \sigma_1^{j,k} \otimes \sigma_2^{j,k} \\
\hline
\mbox{Run 1}            &       |0\rangle                         &      0               &     X               &    \mbox{?}       &         \XBox       \\
\mbox{Run 2}            &  (|0\rangle + |1\rangle)/\sqrt{2}       &      0               &     X               &         0         &         \CheckedBox \\
\mbox{Run 3}            &       |1\rangle                         &      1               &     Z               &         1         &         \CheckedBox \\
\mbox{Run 4}            &       |1\rangle                         &      1               &     Z               &         1         &         \CheckedBox \\
\mbox{Run 5}            &  (|0\rangle - |1\rangle)/\sqrt{2}       &      1               &     Z               &     \mbox{?}      &         \XBox       \\
 %|\psi^{-}\rangle & \mathbb{I} \\
  %          |\psi^{+} \rangle &\sigma_z \\
   %         |\phi^{-} \rangle &\sigma_x  \\
    %        |\phi^{+} \rangle & \sigma_y\\
            \hline
%\mbox{Table 1.} 
\end{array}
\nonumber
\end{equation}
We put a question mark to indicate the runs when Bob's basis does not match with that of Alice, and therefore although Bob does get a 
measurement outcome in these cases, the outcome does not have any correlation with the state of Alice. These are exactly the runs (runs 1 and 5) 
for which Alice and Bob remove the entries in their respective binary digit columns (column 3 for Alice, and column 5 for Bob). 
Note that the binary sequence with Alice, after removal of the non-matchings, is 011, which is the same as that with Bob.

But what if the eavesdropper Eve is operating on the quantum channel that carries the qubits from Alice to Bob? Since the set of states that Alice sends 
to Bob contains mutually nonorthogonal states, the no-information leaking theorem (as discussed in Sec. \ref{sec-cloning}) implies that 
Eve will not be able to obtain any information about the sent qubits without disturbing them. Such disturbance will then show up 
in the measurement results of Bob: The binary sequence at Alice and that at Bob will not match exactly. Therefore to find whether 
Eve is operating on the channel (i.e. whether Eve has obtained any information about the sent qubits), Alice 
%and Bob 
now publicly announces some randomly chosen bits of her binary sequence, and Bob publicly declares the bits at exactly those positions in his
binary sequence. If there is no mismatch, Alice and Bob can be sure, within statistical uncertainty, that the \emph{rest} of the 
binary sequences can be used as a shared key. 
And if there is a mismatch, it implies that Eve has some information about the sent qubits. In that case, they need to start afresh.
This is the BB84 protocol,  and the security of this key distribution scheme depends on the validity of the statement of 
the no-information leaking theorem. The theorem is of course valid within the 
purview of quantum mechanics.

The situation is quite different when the protocol is actually realized. 
In that case, 
%However, in realizing such a protocol, 
``Alice'' and ``Bob'' will be using some real quantum channel, which will produce some noise on its own (i.e. even without the presence of Eve). But Alice and Bob 
have no way but to consider that all the noise have been generated by Eve. Moreover, it may be the case that Eve is always operating on the channel, and 
therefore there is always a mismatch in the binary sequences at Alice and Bob. Quantum mechanics is able to provide security for keys generated in such scenarios
as well. See \cite{reviewcrypto}, and references therein for further details.

%They then look back 
%and then they keep those states in which their choice of bases are same. In this way, finally they share perfect keys. 

%If an evesdropper can have access on the quantum channel by which Alice sends the state, the security of the protocol can be reduced. Quantum no-cloning theorem 
%\cite{nocloning} states that two nonorthogonal quantum states cannot be copied without violating unitarity of quantum mechanics. Therefore, no-cloning guarantees 
%that eavesdropper cannot get full information without getting detected. However, the approximate cloning \cite{approximate} is possible, and hence it provides the 
%bounds on security of quantum key distribution. 

As mentioned earlier, the BB84 protocol does not involve any entanglement. 
In 1991, A. Ekert \cite{Ekertcrypto} discovered a key distribution protocol that uses entangled states for sharing secret 
keys. 
%However it turns out that the Ekert 1991 (E91) protocol is equivalent 
%to the BB84 one \cite{reviewcrypto, BBM, curty-and-others}
However, one can show that entanglement-based
protocols -- like the Ekert 1991 (E91) protocol -- are 
%known to be 
equivalent, in principle, to protocols
that employ quantum channels but do not require entanglement -- like the BB84 protocol \cite{BBM, reviewcrypto, curty-and-others, device-independent}. 
This is 
the status for the case when we are considering security of key distribution protocols involving  a single sender 
and a single receiver. The situation however changes in networks, as we will mention later. 

%Both the protocols are equivalent in the most of the attacks of evesdropper \cite{reviewcrypto}.  

\subsection{Quantum Cryptography in Networks}

There are several interesting quantum cryptographic schemes in the multi-user domain (i.e. beyond the single-sender single-receiver scenario). These 
include secret sharing \cite{hillery1999a,clevea}, and the somewhat different Byzantine agreement problem \cite{Byzantine}. 
Below we briefly consider the secret sharing protocols and some of its security aspects.

\subsubsection{Secret Sharing}

Secret sharing 
%The quantum communication task that we investigate, is 
%known as secret sharing 
\cite{hillery1999a} (cf. \cite{clevea}) is a communication task in which 
a sender, Alice,  wants to send
a (classical) message to several 
%two
recipients 
(called ``Bobs'' -- \(B_1, B_2, \cdots\)),
so that each of the Bobs
individually knows nothing about the message, but 
can recover its content once they cooperate.
For simplicity, let us consider the case of two receivers.
For transmitting a binary message string $\{a_i\}$, Alice can 
send 
a random sequence of bits, $\{b_{1,i}\}$,
to $B_1$, and 
the sequence $\{b_{2,i}\}=\{a_i + b_{1,i}\}$ to
$B_2$, where again we are considering bitwise addition modulo $2$.
Since
$a_i = b_{1,i} +  b_{2,i}$, we are assured that the Bobs can recover the message if they cooperate, and yet
none of them can learn anything about  the message of Alice on his own, since the sequences $\{b_{1,i}\}$,
$\{b_{2,i}\}$ are completely random. 
In addition to the requirement that the receivers must not learn anything about the message unless they cooperate, 
it is also demanded that no third (actually fourth (!), assuming that  there are two receivers) party learns about the message sent by Alice. 
%
%
%An important issue is of course security, i.e. distributing the message in a way
%that no third  (actually fourth!) party learns  about it. 
The demands can be met by 
%This can be achieved
using the single-sender single-receiver quantum cryptography protocols described before (e.g. by the BB84 or the E91 schemes).
The procedure goes as follows: Alice  establishes secret random keys, independently, with both Bobs,
and uses them as one-time pads to securely send bits in the way required by secret sharing.
Let us call this the BB\(84^{\otimes 2}\) protocol.
However, as argued in \cite{hillery1999a},  
 a more natural way of using quantum states in secret sharing is
to send entangled states to the Bobs,
and as a result, avoid establishing random keys with each of the Bobs separately,
by combining the quantum and classical
parts of secret sharing in a single protocol. The authors of \cite{hillery1999a} go on to describe a protocol 
that involves sending four bipartite entangled states by Alice to the two Bobs.
%We call the protocol in \cite{hillery1999a} as E4 (s
Since it uses four entangled states, let us call the protocol as the E4 protocol.
%
%
%It may be noted here that
%without the LOCC constraint, 
The secret sharing protocol has been experimentally realized. See e.g. 
\cite{secret-sharing-expt}.

\textbf{Entanglement enhances security in quantum communication:}
As mentioned before, entanglement has
been identified as the essential ingredient in quantum communication
without a security aspect, e.g., in quantum dense
coding and quantum teleportation.
However in secure quantum communication, if we restrict ourselves to the case of a single sender and a single receiver, entanglement-based protocols are equivalent 
to the product state-based ones. Entanglement is
also useful in secure quantum communication, but one has then to go  beyond the bipartite case.

Let us again consider the secret sharing scheme described before, and 
let us begin with the case when the eavesdropper has access to global quantum operations on both the quantum channels -- 
one from Alice to \(B_1\) and another from Alice to \(B_2\).
%Consider first the case when the eavesdropper has access to global quantum operations on both the quantum channels -- one from Alice to \(B_1\) and another
%from Alice to \(B_2\). 
In this case, 
the security analyses of the E4
secret sharing protocol and the single-sender single-receiver
BB84 cryptographic protocol are isomorphic, 
%if one considers that the eavesdropper has access to global quantum operations on both the 
as both protocols
make use of four nonorthogonal states with the same
mutual scalar products. See \cite{hillery1999a} (cf. \cite{secret-sharing-amader+}). 
However such ``global'' eavesdropping attacks are not the most important ones in a cryptographic scenario with separated receivers, as in secret sharing.

Ref. \cite{local-Eve} finds optimal eavesdropping attacks, both for E4 and BB\(84^{\otimes 2}\) protocols, that are \emph{local} -- there are two 
eavesdroppers, one acting with local quantum operations on the \(A \rightarrow B_1\) quantum channel while the other attacking 
with the same class of operations on the \(A \rightarrow B_2\) one, and they communicate among themselves over a (secure!) classical channel. It turns 
out that the entanglement-based E4 protocol provides a higher security threshold than the product state-based BB\(84^{\otimes 2}\) one, in terms of what is 
known as the ``tolerable quantum bit error rate''.

\section{In lieu of a conclusion}

%In place of a conclusion, let us discuss a few  important problems that would be great to solve but are as yet open. 
Being a young subject, there is 
no dearth of open problems in quantum information in general, and in quantum communication in particular.
%, and the selection here is somewhat arbitrary. 
%\begin{enumerate}
A criterion to detect entanglement is not known even in the bipartite regime. 
%Due to the complexity in the multiparty domain, 
Characterization and quantification of entanglement is difficult even for pure  states in the multiparty case. 
%Although 
Many classical and quantum information transmission protocols have been theoretically proposed and several of  them have already 
been realized in the laboratory in different physical systems for the case of a single sender and a single receiver. 
But quantitative understanding is missing in many cases, and and many additivity questions are unanswered. On the experimental front, 
achieving high fidelities and long distances are some of the main difficulties.  
The complexity of the multiparty terrain -- both on the theoretical and 
experimental fronts -- restricts the study of information  transfer in such systems. 
%The problem has been solved for two qubit systems, and for a qubit-qutrit system \cite{qutrit, HHH-PPT}.
%\item \com{what else} 
%\end{enumerate}

However, a lot of progress has been made in the past decade and a half in the understanding of the theoretical aspects of this new science, and 
experiments realizing quantum communication protocols have in the meanwhile metamorphosed from table-top ones to those between nearby towns and even between 
 nearby islands.


\begin{thebibliography}{}
\bibitem{easy-hard} The terms ``easy'' and ``hard'' used here have precise  meanings in complexity theory. See e.g. \cite{classical-crypto-book}.
\bibitem{asymp} The Holevo bound is asymptotically  achievable in the sense that if
the sender Alice is able to wait long enough and send long strings
of the input quantum states, then there exists a
particular encoding and a decoding scheme that asymptotically
attains the bound. Moreover, the encoding consists in collecting
certain long and ``typical'' strings of the input states, and
sending them all at once \cite{babarey, maarey}.
Note however that the capacity defined here is the so-called ``product'' capacity, where we assume that during the encoding phase, we do not use 
entanglement over different uses of the channel. Very recently, it has been shown that using entangled encoding can actually lead to higher, 
as yet by a small amount, capacities \cite{Hastings}.
\bibitem{mixed-input} We will not have ocassion to consider the case when also the input to the machine is mixed.
\end{thebibliography}
\end{document}